\documentclass[12pt,xcolor]{article}

\usepackage{amssymb,amsmath,amsthm}
\usepackage{url}
\usepackage{natbib}
\usepackage{graphicx}
\usepackage[mathscr]{eucal}
\usepackage[OT2,T1]{fontenc}
\usepackage{verbatim}
\usepackage{eurosym}

\newcommand{\bG}{\mathbb{G}}
\newcommand{\nE}{n_E}
\newcommand{\nI}{n_I}
\newcommand{\pEE}{p_{EE}}
\newcommand{\pII}{p_{II}}
\newcommand{\pEI}{p_{EI}}
\newcommand{\pIE}{p_{IE}}

\newcommand{\mE}{\mathcal{E}}
\newcommand{\mI}{\mathcal{I}}
\newcommand{\ve}{\varepsilon}
\newcommand{\tw}{\widetilde{w}}
\newcommand{\tX}{\widetilde{X}}
\newcommand{\btX}{\overline{\widetilde{X}}}

\newcommand{\ptX}{p_{\widetilde{X}}}

\newcommand{\zstare}{z^{\prime}_{\ve}}

\newcommand{\bth}{\boldsymbol{\theta}}

\newcommand{\mc}[1]{\mathcal{#1}}
\newcommand{\mb}[1]{\boldsymbol{#1}}
\newcommand{\mbb}[1]{\mathbb{#1}}

\newcommand{\PP}{\mathbb{P}}

\newcommand{\bTh}{\mb{\Theta}}

\begin{document}

\title{Optimizing the quantity/quality trade-off\\in\\connectome inference\footnote{This work is partially supported by the Research Program in Applied Neuroscience.}}
\author{Carey E.\ Priebe$^1$, Joshua Vogelstein$^1$, Davi Bock$^2$\\$^1$JHU AMS \& $^2$HHMI Janelia}
\date{\today}
\maketitle
\abstract{
We demonstrate a meaningful prospective power analysis for an (admittedly idealized) illustrative connectome inference task.
Modeling neurons as vertices and synapses as edges in a simple random graph model,
we optimize the trade-off between the number of (putative) edges identified and the accuracy of the edge identification procedure.
We conclude that explicit analysis of the quantity/quality trade-off is imperative for optimal neuroscientific experimental design.
In particular, identifying edges faster/more cheaply, but with more error, can yield superior inferential performance.
}

\section*{Introduction}

Statistical inference on graphs begins with modeling graph-valued observations $G=(V,E)$,
where $V=\{1,\cdots,n\}$ is the vertex set and
$E \subset V \times V$ is the edge set (conections between vertices),
via a random graph model $\bG \sim P_{\theta_0} \in \mc{P} = \{P_{\theta}:\theta \in \Theta\}$.
The parameter $\theta$ governs the distribution $P_{\theta}$
over the collection $\mc{G}_n$ of possible graphs on $n$ vertices,
and the parameter set $\Theta$ indexes the distributions in our model.
Inference then proceeds via
estimation or hypothesis testing
regarding the true but unknown parameter value $\theta_0 \in \Theta$.

Statistical inference on connectomes --
graphs representing brain structure --
involves positing a probabilistic model for the connectome
and deriving desirable properties of a statistic (a function of $G$ or $\bG$)
with respect to neuroscientific questions regarding $\theta_0$.

For example, electron microscopy (EM) and magentic resonance (MR) imaging technology
can produce high-resolution connectome data.
In EM,
the connectome is the graph obtained by representing neurons as vertices and synapses as edges.
In MR, the vertices represent voxels or neuroanatomical regions
and the edges represent functional, effective, or structural connectivities.
It is estimated that a human or primate cortical column
contains approximately 100,000 neurons,
each with about 10,000 connections,
yielding approximately one billion synapses.
Thus one could hope to observe a massive graph
and perform inference thereon,
yielding fundamentally important neuroscientific understanding.

However, given the imaging data, one must estimate the graph.
Typically, these brain-graphs are obtained by tedious and time consuming manual annotation.
In \cite{Bock2011}, approximately nine expert-human months were required to find 250 synapses
in EM imagery of the mouse primary visual cortex -- about 1 synapse per expert-human day.
At that rate, it would take nearly 300 million expert-human years to recover the full induced subgraph of a primate cortical column.
It is possible that annotating more quickly -- and more errorfully --
would yield superior statistical inference.
Indeed, regardless of the scale of the connectome, or the imaging technology,
there is an inherent quantity/quality trade-off in statistical connectomics.

In this manuscript,
we present an (admittedly idealized) illustrative setting in which
we optimize the quantity/quality trade-off analytically,
demonstrating that identifying brain-graph edges faster/more cheaply, but with more error,
can yield superior inference in statistical connectomics.
We describe a very simple brain-graph model and a correspondingly simple error model
to explicate how the quantity/quality trade-off impacts the power of a particular hypothesis test.

\section*{Connectomic Motivation}

The connections made by cortical brain cells are anatomically nanoscopic,
yet each cell in the cortex has several centimeters of local anatomical ``wiring'' (\cite{Cortex1998}).
This wiring packs the cortical volume essentially completely.
\cite{Bock2011} recently characterized the {\it in vivo} responses of a group of cells in mouse visual cortex, then imaged
a volume of brain containing the cells using a custom-built high throughput electron microscopy (EM) camera array. Each voxel in the resulting data set occupies
about 4 x 4 x 45 cubic nanometers of brain; the 10 teravoxel volume spans 450 x 350 x 50 cubic micrometers. The imaged volume is of sufficient size and resolution that they
were able to trace the local connectivity of the physiologically characterized cells. One can therefore record what cells in the brain are doing and then trace
their connectivity -- a combination which could enable a new level of understanding of cortical circuits to be achieved.

\section*{Model \& Hypotheses}

Let $\bG$ be an independent edge stochastic block model on $n$ vertices.

Vertices represent neurons.
We denote by $\mE$ the collection of $\nE$ excitatory neurons
and by $\mI$ the collection of $\nI$ inhibitory neurons.
Thus the vertex set $V$ is decomposed as the disjoint union $V = \mE \cup \mI$.
Let $n = |V| = |\mE| + |\mI| = \nE + \nI$
with $\nE = \lambda n$ and $\nI = (1-\lambda) n$ for some $\lambda \in (0,1)$.

Edges represent synapses.
We consider loopy graphs -- a neuron may connect to itself.
We consider the undirected edge case for simplicity;
the directed and multigraph cases follow {\em mutatis mutandis}.

The block model structure is such that
\begin{eqnarray*}
 P[u \sim v] &=& \pEE \mbox{ ~ for ~ } u,v \in \mE, \\
 P[u \sim v] &=& \pII \mbox{ ~ for ~ } u,v \in \mI, \\
 P[u \sim v] &=& \pEI = \pIE \mbox{ ~ otherwise}.
\end{eqnarray*}
That is,
the probability that two excitatory neurons connect to one another
is given by random graph model parameter $\pEE$,
the probability that two inhibitory neurons connect to one another
is given by $\pII$,
and
the probability that an excitatory neuron connects to an inhibitory neuron
is given by $\pEI=\pIE$ (necessarily equal if and only if we are considering the undirected case).

We assume that the excitatory-excitatory connection rate $\pEE$
and the inhibitory-inhibitory connection rate $\pII$
are equal.
We propose to test the hypothesis that this common rate
 $\pEE = \pII$ is equal to the excitatory-inhibitory connection rate $\pEI$:
\begin{eqnarray*}
 H_0:&& \pEE = \pII = \pEI \\
 vs.&& \\
 H_A:&& \pEE = \pII < \pEI ~ .
\end{eqnarray*}
The available data are an observed collection of {\em putative} or {\em errorful} edges.

\section*{Data}

Ideally, we would observe the entire induced subgraph $G=\Omega(V;G^*)$
for our imaged volume, where $G^*$ is the entire brain graph (connectome).
Practically, since identifying edges is expensive,
for large $n$ we will observe a subgraph -- a subset of edges.
For $i=1,\cdots,z$, we define the random variable $X_i$ representing a perfect edge observation via the ``tracing algorithm'' given by
\begin{enumerate}
\item[(1)] a neuron: choose a vertex $v_i$ uniformly at random from $V$.
\item[(2)] a synapse: choose an edge $v_i \sim \cdot$ uniformly at random from among edges incident to $v_i$.
\item[(3)] the post-synaptic neuron: identify vertex $w_i$ for $v_i \sim w_i$.
\item[(4)] the nature of the synapse: $X_i = I\{ v_i,w_i \in \mE \mbox{ ~ or ~ } v_i,w_i \in \mI \}$.
\item[$\bullet$] If $w_i$ is not in our imaged volume,
or if the edge = axon-synapse-dendrite goes outside our imaged volume so that we cannot trace it, we try again (with the same $i$).
\item[$\bullet$] If $v_i \sim w_i$ is by chance previously identified, we try again (with the same $i$).
\end{enumerate}
However, the algorithm above requires {\em perfect} tracing.
Unfortunately, perfect edge observations are expensive, even for a subset of edges.
Instead, with error probability $\ve \in [0,1]$ we fail to correctly identify $w_i$ and instead
identify a randomly chosen vertex $\tw_i$ in Step (3) above, yielding
\begin{enumerate}
\item[($3^{\prime}$)] identify $\tw_i$ for $v_i \sim \tw_i$; with probability $(1-\ve)$ $\tw_i=w_i$, otherwise $\tw_i$ is random.
\item[($4^{\prime}$)] $\tX_i = I\{ v_i,\tw_i \in \mE \mbox{ ~ or ~ } v_i,\tw_i \in \mI \}$.
\end{enumerate}
Thus $\tX_i=1$ if the $i^{th}$ (putative) edge is either
an excitatory-excitatory connection or an inhibitory-inhibitory connection,
and $\tX_i=0$ if the $i^{th}$ (putative) edge is an excitatory-inhibitory connection.
The edge tracing algorithm generates $z$ such $\ve$-errorful edges,
yielding an errorful subgraph observation model.

\section*{Trade-off}

Presumably, the expense of edge tracing is increasing in putative edge count $z$ for fixed edge tracing error $\ve$ and decreasing in $\ve$ for fixed $z$.
For fixed resources (imaging resolution and/or manual or automatic edge tracing resources)
the trade-off of interest is $z$ vs.\ $\ve$ -- quantity vs.\ quality.
The number of $\ve$-errorful edges that will be traced, $z=h(\ve)$, is an increasing function of $\ve$.
We derive below the optimal operating point of the quantity/quality trade-off in a particular connectome inference task.

Of course, committing more resources (higher-resolution imaging and/or additional edge tracing resources)
should yield larger $z$ for the same $\ve$ or smaller $\ve$ for the same $z$;
a prospective power cost/benefit analysis can be performed to aid in the decision regarding commitment of resources.
Still, the quantity/quality trade-off is an essential component of any such cost/benefit analysis,
since one would want to consider the optimal quantity/quality operating point for each level of resource commitment.

\section*{Inference}

We derive the expression for
\begin{eqnarray*}
P[\tX_i = 1] = \ptX &=& \ptX(n,\lambda,\pEE,\pEI,\ve)\\
&=& (1-\ve) \left( \frac{\lambda \pEE n_E}{\pEE n_E + \pEI n_I}
                 + \frac{(1-\lambda) \pII n_I}{\pII n_I + \pIE n_E} \right)
  + \ve (2\lambda^2-2\lambda+1).
\end{eqnarray*}
Note that under $H_0$ the value of $\ptX(n,\lambda,\pEE,\pEI,\ve)$ is independent of the value of $\pEE=\pEI$,
and that $\ptX$ is {\em smaller} under the alternative hypothesis $\pEE < \pEI$ than under the null.
Since we have (approximately) independent random variables $\tX_i \sim Bernoulli(\ptX)$,
we reject for small values of the test statistic
$\btX_z = \frac{1}{z}\sum_{i=1}^z \tX_i$
based on having observed $z$ errorful edges.
Assuming independent errors, this test is uniformly most powerful (UMP).
Applying the central limit theorem under both $H_0$ and $H_A$ yields a large $n$ large $z$ normal approximation for the power
of the level $\alpha$ test,
\begin{eqnarray*}
P[\btX_z < c_{\alpha}|H_A] = \beta_{z,\ve} &=& \beta_{z,\ve}(n,\lambda,\pEE,\pEI;\alpha) \\
&=& \Phi\left( \frac{\ptX^0(1-\ptX^0)\Phi^{-1}(\alpha) + \sqrt{z}(\ptX^0-\ptX^A)}{\ptX^A(1-\ptX^A)} \right),
\end{eqnarray*}
where $\ptX^0$ and $\ptX^A$ denote the value of the Bernoulli parameter $\ptX(n,\lambda,\pEE,\pEI,\ve)$ given above
under $H_0$ (whatever be the value of $\pEE=\pEI$) and
under $H_A$ (for specific values of $\pEE < \pEI$), respectively.

Perfect edge tracing ($\ve=0$) for $z$ edges yields power $\beta_{z,0} > \alpha$.
Errorful edge tracing ($\ve>0$) for $z$ putative edges yields power $\beta_{z,\ve} < \beta_{z,0}$.
As expected, more error yields less power for fixed putative edge count $z$:
$\ve_1 < \ve_2$ implies $\beta_{z,\ve_1} > \beta_{z,\ve_2}$
and
$\beta_{z,1} = \alpha$ for any $z$.
Furthermore, more edges yields more power for fixed edge tracing error rate $\ve$:
$z_1 > z_2$ implies $\beta_{z_1,\ve} > \beta_{z_2,\ve}$ for any $\ve$.
However, we can identify equivalent sample size $\zstare$ such that $\beta_{\zstare,\ve} \approx \beta_{z,0}$.
Thus, if errorful edge tracing is sufficiently less expensive so that we can trace more than $\zstare$ errorful edges
 compared to just $z$ perfect edges
(which is plausible since
 perfect edge tracing is expensive while
 errorful edge tracing should be less so)
then inferential performance based on an errorful subgraph will be superior to inferential performance based on a perfect subgraph.
This suggests that we may benefit from optimizing the quantity/quality trade-off with respect to power for fixed resources.

\section*{Example}

A mouse cortical column
(the existence of which is admittedly the subject of neuroscientific debate;
we proceed with an illustrative example regardless)
has approximately 10,000 neurons.
With parameter values
$n=10000, \lambda=0.9, \pEE=\pII=0.1, \pEI=0.2$
for the random graph model $\bG$,
we expect the graph to have roughly 6 million edges total in the induced subgraph.
High-accuracy manual edge tracing results in approximately one edge per expert per day.
For these parameter values, testing at level $\alpha=0.05$ yields
\begin{eqnarray*}
 \beta_{50,0} & \approx & 0.429,\\
 \beta_{50,0.5} & \approx & 0.196,\\
 \beta_{250,0.5} & \approx & 0.488.
\end{eqnarray*}
Thus our prospective power analysis demonstrates that
less expensive errorful edge tracing can be inferentially superior to more expensive perfect edge tracing:
if we can trace $z=50$ edges perfectly ($\ve=0$)
we obtain power $\beta_{50,0} \approx 0.429$
(compared to degraded power $\beta_{50,0.5} \approx 0.196$
with the same number of (putative) edges ($z=50$) and 50\% edge tracing error ($\ve=0.5)$),
while
if we can trace $z=250$ putative edges with 50\% edge tracing error ($\ve=0.5)$
we obtain significantly improved power $\beta_{250,0.5} \approx 0.488 > \beta_{50,0} \approx 0.429$.
The equivalent sample size for this example is $z^{\prime}_{0.5} = 178$, so that $\beta_{178,0.5} \approx \beta_{50,0} \approx 0.429$;
thus tracing more than 178 50\%-errorful putative edges
yields higher power than that obtained with 50 errorless edges.

Extending this example, we assume that $z=h(\ve)$.
That is, the number of (errorful) putative edges that we can trace with edge tracing error $\ve$
is given by some (increasing) function $h$ of $\ve$.
Thus the power $\beta(\ve)$ obtained when using the edge tracing algorithm
engineered to produce $z=h(\ve)$ putative edges with edge tracing error $\ve$
is given by
$$\beta(\ve) = \Phi(g(\ve))$$
where
$$g(\ve) = \frac{\ptX^0(\ve)(1-\ptX^0(\ve))\Phi^{-1}(\alpha) + h(\ve)^{1/2}(\ptX^0(\ve)-\ptX^A(\ve))}{\ptX^A(\ve)(1-\ptX^A(\ve))}.$$
Assuming that $h$ is differentiable with respect to $\ve$ on $[0,1)$,
we obtain
$$\frac{\partial \beta}{\partial \ve} = \phi(g(\ve))g'(\ve).$$
Then we evaluate the sign of
$\frac{\partial \beta}{\partial \ve}|_{\ve=\ve_0}$
at the current edge tracing algorithm operating point $\ve_0$;
$\frac{\partial \beta}{\partial \ve}|_{\ve=\ve_0} > 0$
implies less expensive more errorful (larger $\ve$) edge tracing (resulting in larger $z$) will yield increased power, while
$\frac{\partial \beta}{\partial \ve}|_{\ve=\ve_0} < 0$
implies that inference will improve with more accurate but more expensive edge tracing (resulting in fewer putative edges).
Finding $\ve^{\star}$ such that
$\frac{\partial \beta}{\partial \ve}|_{\ve=\ve^{\star}} = 0$
will (after checking appropriate side conditions) yield optimal power $\beta^{\star}=\beta(\ve^{\star})$.
To continue with our example,
we consider for illustration
$$z=h(\ve)=50+\frac{200}{\sin(\pi/4)}\sin(\ve \pi/2),$$
designed to give
$h(0)=50$, $\beta(0) \approx 0.429$
and
$h(1/2)=250$, $\beta(1/2) \approx 0.488$
for consistency with our running example.
This $h$ suggests that 50 expert days yields $z=50$ at $\ve=0$ and $z=250$ at $\ve=0.5$;
investigation into the precise character of an appropriate $h$ will be a necessary.
For the specified $h$ in our example we calculate the optimal operating point for the edge tracing algorithm,
obtaining $\ve^{\star} \approx 0.247$
and resulting in $h(\ve^{\star}) \approx 157$ and $\beta(\ve^{\star}) \approx 0.599$.
Thus optimizing the quantity/quality trade-off has yielded an improvement in power of almost 40\%.
We should engineer our edge tracing to operate at error rate $\ve^{\star} \approx 0.247$.

\begin{figure}
\includegraphics{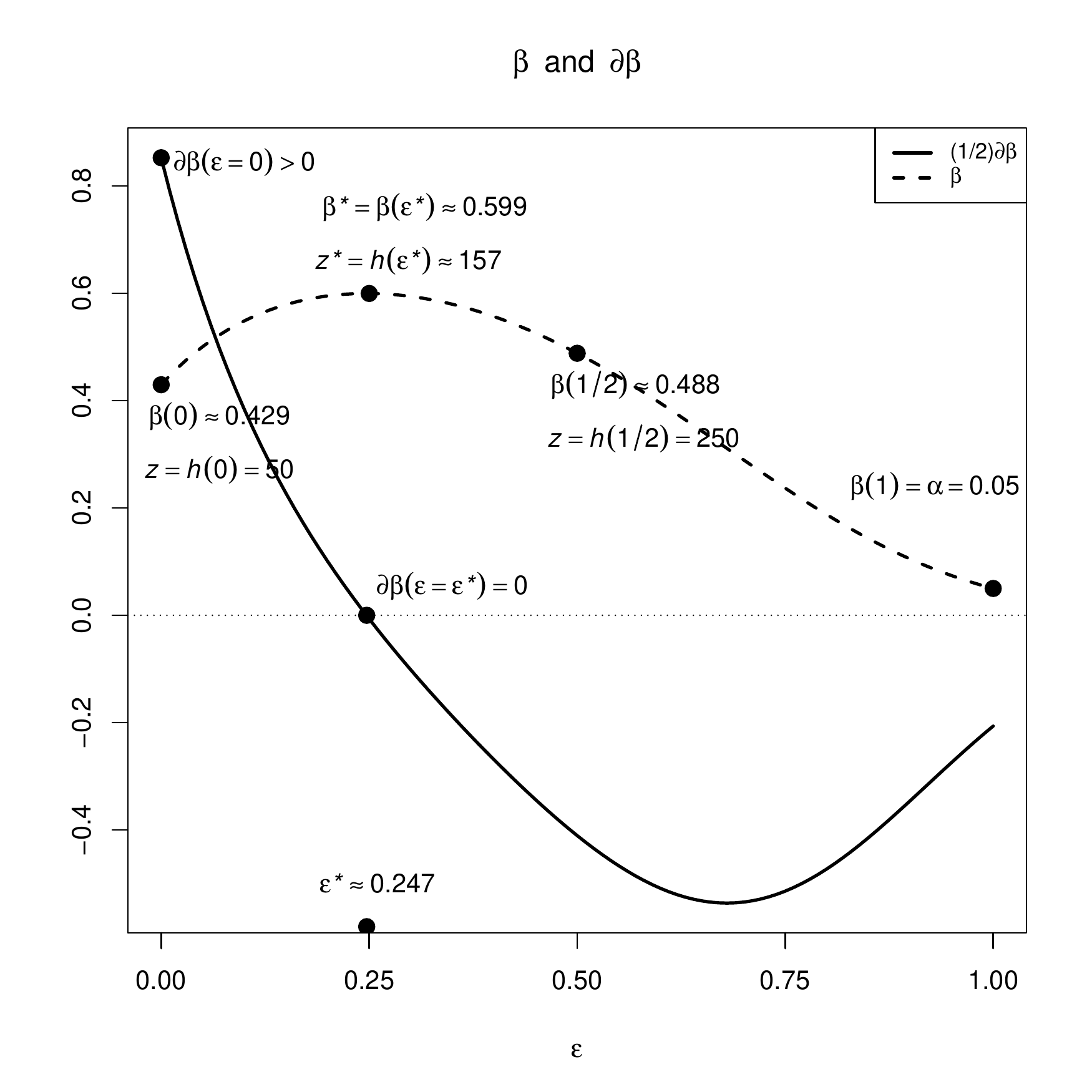}
\caption{\label{betafig}
Power $\beta$ and its derivative $\frac{\partial \beta}{\partial \ve}$
as functions of the edge tracing error rate $\ve$ for our example scenario
(see text for details).
(We plot $(\frac{1}{2})\frac{\partial \beta}{\partial \ve}(\ve)$
so that the two curves are on approximately the same scale
and can productively be presented on the same plot.)
}
\end{figure}

A summary of this example is presented in Figure \ref{betafig}. 

\section*{Discussion}

We conclude that we can indeed do a meaningful prospective power analysis
for this connectome inference task, 
and that analysis of the quantity/quality trade-off between
error in edge tracing and the number of putative edges traced
is imperative for optimal neuroscientific experimental design.

The significance of our ``admittedly idealized'' illustrative setting is a simple version of a general question of scientific interest: how does connectivity probability depend on the neurons in question?  Real scientific interest lies in more elaborate graph models and hypotheses -- $K>2$ kinds of cells and $K^2$ connection probabilities, or even an unknown number of cell types.  The method described here can be generalized to these more realistic settings -- some maintaining analytic tractability, but many realistic complex generalizations will of course require us to resort to numerical approximation methods.
In hypothesis testing, one wishes to collect data in such a way
so as to maximize the probability of rejecting the null hypothesis given that it is false.
Often, the data collector is limited to only experimental intuition
in making the quantity/quality decision.
In some cases, however, one can turn to statistical connectomics to
shed light on the quantitative trade-offs one expects with regard to a particular statistical inference question.
Specifically, we have demonstrated that one can approximate the optimal operating point for the (errorful) edge tracing algorithm.


The above example uses statistical connectomics
to address an important decision in neuroscientific data collection and analysis.
While the results presented apply to a special case, general lessons can be learned.

In particular, we see that
explicitly modeling the quantity/quality trade-off can yield significant inferential advantages.
Note that the optimal operating point depends heavily on specific model assumptions;
thus, any conclusions from such a prospective analysis are subject to the adequacy of those assumptions.
We emphasize that this analysis is fundamentally a function of the particular inference task.
Although we have outlined analytical results for one specific
(i) inference task,
(ii) graph model,
(iii) error model, and
(iv) quantity/quality trade-off function,
each of these components must be customized for the neuroscientific question at hand.

The example results presented herein depend on knowing the quantity/quality function $z=h(\varepsilon)$.
In general, this function will not be known, but it can be estimated.
Specifically, consider the scenario of manual annotation of EM data.
The performance of trained edge tracers operating so as to target various putative edge counts can be calibrated against a ``gold'' standard --
derived, perhaps, using independent, complementary imaging methods --
providing an estimate of $h$.
As the size of connectome data sets continues to increase,
the number of manual annotators required to estimate massive connectomes gets impractically large.
Therefore, we will rely on machine vision algorithms to annotate the data (cf.\ the \cite{OpenConnectomeProject}).
The quantity/quality trade-off applies to such algorithms as surely as it applies to manual annotation,
and the quantity/quality function $h$ will need to be estimated.

The implications of
optimizing the quantity/quality trade-off in connectome inference
are potentially substantial in light of the recent global investment in connectome science.
For example,
the USA National Institute of Health (NIH) has budgeted over \$30 million
to the \cite{HumanConnectomeProject},
which aims to collect and analyze human magnetic resonance (MR) connectomes.
Similarly, the European Union (EU) is potentially granting up to
\EUR{1} billion
to the \cite{HumanBrainProject}.
Understanding and exploiting the quantity/quality trade-off in connectome inference
will be essential to the efficient use of the available resources.


\bibliography{refs}

\begin{thebibliography}{5}
\providecommand{\natexlab}[1]{#1}
\providecommand{\url}[1]{\texttt{#1}}
\expandafter\ifx\csname urlstyle\endcsname\relax
  \providecommand{\doi}[1]{doi: #1}\else
  \providecommand{\doi}{doi: \begingroup \urlstyle{rm}\Url}\fi

\bibitem[Bock et~al.(2011)Bock, Lee, Kerlin, Andermann, Hood, Wetzel,
  Yurgenson, Soucy, Kim, and Reid]{Bock2011}
Davi~D. Bock, Wei-Chung~A. Lee, Aaron~M. Kerlin, Mark~L. Andermann, Greg Hood,
  Arthur~W. Wetzel, Sergey Yurgenson, Edward~R. Soucy, Hyon~S. Kim, and R.~Clay
  Reid.
\newblock {Network anatomy and in vivo physiology of visual cortical neurons}.
\newblock \emph{Nature}, 471\penalty0 (7337):\penalty0 177--182, March 2011.

\bibitem[Braitenberg and Sch{\"u}z(1998)]{Cortex1998}
V~Braitenberg and A~Sch{\"u}z.
\newblock \emph{Cortex: Statistics and Geometry of Neuronal Connectivity}.
\newblock Springer, Berlin, Germany, 1998.

\bibitem[Human Brain Project()]{HumanBrainProject}
Human Brain Project.
\newblock URL \url{http://www.thehumanbrainproject.com/}.

\bibitem[Human Connectome Project()]{HumanConnectomeProject}
Human Connectome Project.
\newblock URL \url{http://www.humanconnectomeproject.org/}.

\bibitem[Open Connectome Project()]{OpenConnectomeProject}
Open Connectome Project.
\newblock URL \url{http://www.openconnectomeproject.org/}.

\end{thebibliography}

\end{document}